\documentclass[twoside]{article}
\usepackage[a4paper]{geometry}
\usepackage[utf8x]{inputenc}
\usepackage[T1]{fontenc}
\usepackage{lmodern}
\usepackage{fRReE}
\usepackage{url}
\usepackage{hyperref}
\usepackage{graphicx}
\usepackage{float} 

\usepackage{listings}

\RRlogo{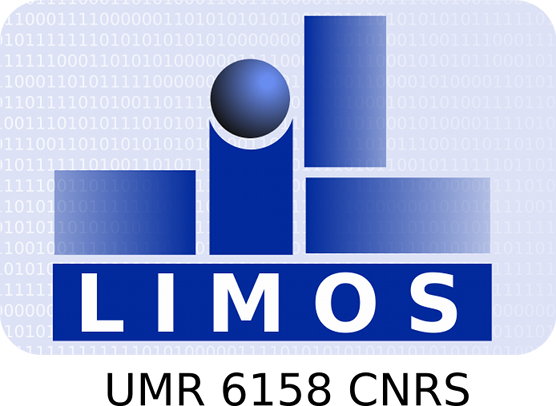}

\RRrc{limos}
\RRrcname{LIMOS - UMR CNRS 6158}
\RRrcaddr{
Campus des C\'ezeaux\\
B\^atiment ISIMA\\
BP 10125 - 63173 Aubi\`ere Cedex\\
France}
\RRrcwebsite{\url{http://limos.isima.fr/}}

\RRdate{July 2012}

\RRauthor{
Jonathan \textsc{Passerat-Palmbach}\RRaffiliation{This author's PhD is partially funded by the Auvergne Regional Council and the FEDER}
\RRaffiliation[sfn0]{ISIMA, Institut Sup\'erieur d'Informatique, de Mod\'elisation et de ses Appplications, BP 10125, F-63173 AUBIERE}
\RRaffiliation[sfn1]{Clermont Universit\'e, Universit\'e Blaise Pascal, LIMOS, BP 10448, F-63000 CLERMONT-FERRAND}
\RRaffiliation[sfn2]{Clermont Universit\'e, Universit\'e Blaise Pascal, BP 10448, F-63000 CLERMONT-FERRAND}
\RRaffiliation[sfn3]{CNRS, UMR 6158, LIMOS, F-63173 AUBIERE}
   \and
\\David R.C. \textsc{Hill}
\RRaffiliationref{sfn0}
\RRaffiliationref{sfn1}
\RRaffiliationref{sfn2}
\RRaffiliationref{sfn3}
}

\RRauthorhead{Passerat-Palmbach et. al}
\RRtitle{How to Correctly Deal With Pseudorandom Numbers in Manycore Environments - Application to GPU programming with Shoverand}
\RRsource{Proceedings of the IEEE High Performance Computing and Simulation conference 2012}
\RRtitlehead{HPCS 2012}

\RRcopyright{\textcopyright 2012 IEEE}
\RRoriginalpages{25-31}
\RRadditionalinformation{http://dx.doi.org/10.1109/HPCSim.2012.6266887\\\textit{(tutorial paper)}}

\RRabstract{
Stochastic simulations are often sensitive to the source of randomness
that characterizes the statistical quality of their results. Consequently, we
need highly reliable Random Number Generators (RNGs) to feed such applications.
Recent developments try to shrink the computation time by relying more and more
General Purpose Graphics Processing Units (GP-GPUs) to speed-up stochastic
simulations. Such devices bring new parallelization possibilities, but they
also introduce new programming difficulties. Since RNGs are at the base of any
stochastic simulation, they also need to be ported to GP-GPU. There is still a
lack of well-designed implementations of quality-proven RNGs on GP-GPU
platforms. In this paper, we introduce ShoveRand, a framework defining common
rules to generate random numbers uniformly on GP-GPU. Our framework is designed
to cope with any GPU-enabled development platform and to expose a
straightforward interface to users. We also provide an existing RNG
implementation with this framework to demonstrate its efficiency in both
development and ease of use.

}
\RRkeyword{pseudorandom numbers, manycore, GP-GPU, High Performance Computing, Shoverand, stochastic simulation}

\begin{document}
\makeRR

\section{Introduction}
More than 50\% of High Performance Computing (HPC) applications now require the
use of Random Number Generators (RNGs). We have had RNGs adapted to intensive parallel computing
at our disposal for more than a decade . But even though the
subject is well mastered by specialists like Pierre L’Ecuyer or Makoto
Matsumoto, this is not the case for many colleagues who are acknowledged as 
specialists in other domains but are not aware of the recent parallelization
techniques than can be used with modern generators. Even if we now also have
access to true random numbers with quantum devices, they are still clumsy to
produce in parallel with satisfying rates and costs. In addition, such devices
are still subject to partial breakdowns and even the latest ones do have some
very little biases that have to be corrected (such processing is sometimes
included inside the device). In addition, stochastic experiments have to be
reproducible, not only for debugging purposes but also for variance reduction
techniques, for sensitivity analysis and for many other statistical techniques
\cite{Kleijnen1986}. This implies storing the sequences (in order to be able to
reproduce exactly the same sequence several times). This last point leads to
the fact that we cannot provide enough "true" numbers for many High Performance
Computing applications such as parallel Monte Carlo for nuclear medicine
\cite{Lazaro.etal.2005, ElBitar.etal.2006}. Consequently, software random number
generation remains the prevailing method for HPC, and specialists have been
warning us for years to be particularly careful when dealing with parallel
stochastic simulations \cite{DeMatteis.Pagnutti1995, Hellekalek1998, Pawlikowski2003}.

Recent developments try to shrink computation time by relying more and more on
General Purpose Graphics Processing Units (GP-GPUs) to speed-up stochastic
simulations. Such devices bring new parallelization possibilities, but they
also introduce new programming difficulties. Since the introduction of Tesla
boards, Nvidia, ATI and other manufacturers of GP-GPUs have changed the way we
use our high computing performance resources. Since 2010, we have seen that the
top supercomputers are now often hybrid. Since RNGs are at the base of any
stochastic simulation, they also need to be ported to GP-GPU. There is still a
lack of well-designed implementations of quality-proven RNGs on GP-GPU
platforms. Consequently, we also need a survey of the current Pseudo Random
Numbers Generators (PRNG) available on GPU. We will discuss the recent Mersenne
Twister for Graphics Processors (MTGP) that was initially released in 2010, but
also more recent generators with cryptographic inspiration that have been
presented lately at the annual supercomputing conference. 

In this paper we intend to present the good practices when dealing with
pseudorandom streams in parallel. After explaining the kind of problems we may
encounter, the state of the art in terms of RNGs, testing suite and
parallelization techniques is presented. Then, GPU considerations and hybrid
computing constraints are exposed. Next, the Shoverand framework that
facilitates a safe usage of pseudorandom streams for modern GPU hardware is
presented, and we also show how to integrate your preferred PRNG in Shoverand if
your RNG of choice is not already included in the framework.

\section{Using Parallel Random Streams}
\subsection{Parallel stochastic simulation}
Parallel and Distributed Simulation (PDS) is an extensive research domain where
effective solutions have been developed. Deterministic communications protocols
for synchronous simulation and asynchronous simulation have been studied to
avoid deadlocks and to preserve the causality and determinism principles. When
dealing with stochastic simulations, random numbers should be generated in
parallel in order for each Processing Element (PE) to be able to autonomously get
its own independent random number stream. If such autonomy is not guaranteed,
the parallelism is affected \cite{Coddington.Ko1998}. 

\subsection{State of the art in terms of sequential generators}
The most famous sequential generator currently available is Mersenne Twister (MT
hereafter) \cite{Matsumoto.Nishimura1998}. Since the initial proposition at the
end of the nineties, a whole family of MT generators has been proposed. SFMT, a
member of this family, is an SIMD-oriented version of the original MT generator
\cite{Saito.Matsumoto2008}. SFMT proposed the following improvements: speed
(twice as fast as MT), a better equidistribution, a quicker recovery from bad
initialization (zero-excess in the initial state) and even an increased 
period length (ranging from $2^{607} - 1$ to $2^{216091} - 1$). A GP-GPU version of MT
(named MTGP) was also proposed by Saito and Matsumoto; it comes with
companion software for parallelization (MTGPDC) \cite{Saito.Matsumoto2012}.
Based on similar principles (Generalized Feedback Shift Register), Panneton, in
collaboration with L’Ecuyer and Matsumoto, proposed the WELL generators
with even better statistical properties \cite{Panneton.etal.2006}.

Pierre L’Ecuyer suggests that multiple recurrence generators (MRGs) with much
smaller periods (above $2^{100}$ but under $2^{200}$) like MRG32k3a
\cite{LEcuyer.etal.2002} can suffice for modern applications. This generator
comes with very interesting statistical properties and is easy to parallelize.
Smaller periods also authorize fast mathematical parallelization techniques.
The MT family now proposes TinyMT which can benefit of this point. In 2011,
Salmon et al recently introduced statistically sound counter based pseudorandom
generators with relatively small periods \cite{Salmon.etal.2011}. These proposals led
to very interesting implementations on GP-GPUs.

With this large set of ‘good’ sequential RNGs, the main question is: how can we
make a safe RNG repartition in order to keep, on the one hand, efficiency, and
on the other hand a sound statistical quality of the simulation in order to
obtain reliable results.

\subsection{Partitionning techniques}
Assigning random sequences to parallel processing elements (PEs) can be done
with one of the two following approaches. The first one proposes to partition a
unique random stream, whereas the second approach deals with multiple
independent streams. In this case, independent streams are obtained by
parameterizing a family of generators. 

When dealing with the partitioning of a unique random source we find the
following variants in literature. The Central Server approach runs a single
PRNG and provides on demand pseudorandom numbers to different PEs. We can also
cite Boolean cellular automata since they have been considered to generate
parallel pseudorandom numbers, but we are not aware of any recent high performance
stochastic simulations using this technique. Another technique is known as the
Sequence Splitting method, sometimes named Blocking or Regular Spacing. The
principle behind those names is simple: the main sequence is split into ‘N’
non-overlapping contiguous blocks. Until recently, the computing of a jump in
the sequence was unavailable for modern generators with linear recurrences
modulo 2 and huge periods (such as the families of MT \& WELL generators). This
problem was solved in \cite{Haramoto.etal.2008}, but despite its relative
efficiency, it is still considered as slow by specialists (compared to what can
be obtained for MRG32k3a with a smaller period). The Indexed Sequence, or
Random Spacing Technique, is another technique that consists in initializing
the same generator with different random statuses. The last known technique to
partition a unique stream is known as the Leap Frog Technique. With this
method, random numbers are distributed to processing elements like a deck of
cards dealt to card players. 

When a single stream is partitioned into many different substreams, we talk
about Parameterization. This technique issues many instances from the same 
family of generators. Independent generators, with
different parameters, are used for each processing element (PE). For the
Mersenne Twister family this technique is named Dynamic Creator (DC) and
generates by parameterization highly independent Mersenne Twister generators.
Our best mathematicians state that PNRGs based on linear recurrences defined by
matrices with characteristic polynomials relatively prime to each other are
supposed to be highly independent. In addition to the original MT generator,
TinyMT and MTGP for GP-GPUs benefit from the same approach. Even though it is
considered safe according to the scientific community, we recommend testing the
resulting generators. In case of failure we can give feedback to authors and
this was recently done for MTGP \cite{Passerat-Palmbach.etal.2010}.

The interested reader can find the details of the different approaches in the
following surveys \cite{Hill2010, Hill.etal.2012}, the latter being the most
complete reference (a full journal paper with the latest updates), with
discussions around advantages and drawbacks of each technique.

\subsection{Testing sequential and paralell RNGs}
The previous ‘fine’ generators were found reliable thanks to thorough
statistical and empirical testing. Testing techniques were initially
proposed by Knuth and the first testing software that became famous in the
nineties was the DieHard testing suite proposed by Marsaglia. Another famous
Statistical Testing Suite is proposed by the National Institute for Standards
and Technology (NIST STS) and is particularly appreciated when cryptographic
qualities are required. \cite{Brown.etal.2006} developed the DieHarder suite as an
update of Marsaglia’s work. Rütti and Troyer also presented a testing suite
with Petersen \cite{Rutti.etal.2004}. But currently, the most complete collection of
utilities for the empirical and statistical testing of uniform random number
generators is in our opinion TestU01. This software library proposed by
\cite{LEcuyer.Simard2007} is now the reference for most scientists of the
domain. 

For parallel testing, Srinivasan and Mascagni gave interesting advice
\cite{Srinivasan.etal.2003} based on the tests proposed by Mascagni in the Scalable
library for Pseudorandom Number Generation (SPRNG) in 1997. Approximately at
the same time, \cite{Coddington.Ko1998} proposed a set of techniques for
empirical testing of parallel random number generators. We have to be aware
that we do not have at our disposal mathematical theorems or techniques to
explicitly give a proof of independence between generators or between generated
parallel random streams. To avoid long-range correlations we can have a look at
interesting propositions made by \cite{DeMatteis.Pagnutti1988, DeMatteis.Pagnutti1995}.

\subsection{Software aid for the distribution of random streams}
The reference software library dealing with parallelization of pseudorandom
numbers is SPRNG (Scalable library for Pseudorandom Number Generation)
\cite{Mascagni1999}. As stated before, SPRNG also proposes a small test suite. The
Dynamic Creator discussed above and proposed initially by Matsumoto and
Nishimura is a software aid to provide mutually independent Mersenne twister
generators for parallel computing, which was published two years later
\cite{Matsumoto.Nishimura2000}. L’Ecuyer and his team proposed a package able
to produce many long streams and substreams in C, C++, Java, and FORTRAN
\cite{LEcuyer.etal.2002} and also in R in a later version. Coddington and Newell
proposed the JaPara library (a Java Parallel Random Number Library) for
High-Performance Computing \cite{Coddington.Newell2004}. If we consider Java,
our advice would be to use the SSJ package, still from L’Ecuyer’s team. It
provides the basic structures for handling multiple streams, with efficient
methods to move streams around \cite{LEcuyer.Buist2005}.

We also proposed higher level software tools to help in the distribution of
stochastic simulations on local clusters and institutional computing Grids. The
DistMe toolkit \cite{Reuillon.etal.2008, Reuillon.etal.2011} and the OpenMole
software based on declarative task delegation \cite{Reuillon.etal.2010} helps
removing the burden of random streams distribution.

In the first part of this paper, we have given a shallow survey of distribution techniques, libraries and tools.
The second part focuses on Shoverand: our library proposal integrating the
guidelines introduced in the previous lines.

\section{Purpose of Shoverand}
\label{purpose_shoverand}
As we have seen in the previous parts of this article, it is very important to
deal carefully with pseudorandom numbers distribution when working in parallel
environments such as GPUs. Still we cannot ask any user that wants to leverage
GPUs' computational power in his stochastic simulations to be
aware of every theoretical consideration that will prevent his simulation
results from being biased. Thus, we introduced Shoverand
\cite{Passerat-Palmbach.etal.2011}, a framework that provides Pseudorandom Number
Generation (PRNG) facilities to CUDA-enabled GPU applications.

Shoverand combines several aspects to ease developments of stochastic-enabled
applications on GPU. First, its API is quite similar to what can be encountered
when using high-level CPU languages like C++ or Java. Second,
Shoverand's main goal is to handle the distribution of
stochastic streams automatically without any intervention from the user.
Finally, our framework also targets PRNG developers: indeed, Shoverand only
integrates third-party PRNGs and focuses on unifying their interface. To do so,
we integrate compile-time constraints that check whether the algorithm meets
our guidelines.

\section{PRNGs embedded in Shoverand}
\label{embedded_prngs}
At the time of writing, Shoverand embeds several PRNGs. All these algorithms
have been selected according to their intrinsic properties. We first consider
their statistical properties in a sequential environment, because a PRNG could
not cope with the requirements of parallel environments if its sequential
version was poor. Consequently, every PRNG wrapped in Shoverand must satisfy
the most stringent testing battery currently available, namely BigCrush from
TestU01 \cite{LEcuyer.Simard2007}. PRNGs that pass all those tests are referred to
as "Crush-resistant" in \cite{Salmon.etal.2011}. While being Crush-resistant cannot
ensure a perfect randomness of the considered pseudorandom stream, it is a
satisfying property that few PRNGs can be proud of.

Additionally, the retained algorithms must support a reliable technique to
distribute numbers in a parallel environment. We have previously surveyed such
techniques in \cite{Hill.etal.2012}, but only some of them can be applied on a GPU
platform \cite{Passerat-Palmbach.etal.2012}. The chosen ones are then ideal
candidates to be ported to GPU, if not available yet, and moreover to be
integrated in Shoverand. We detail hereafter the PRNGs that are currently, or
will soon be, embedded in Shoverand.

\subsection{MRG32k3a}

Introduced by Pierre L'Ecuyer in \cite{LEcuyer1999}, MRG32k3a is
particularly suited to parallelization among small computational elements such
as threads thanks to its intrinsic properties. This PRNG's
lightweight data structure only stores 6 integers to handle its state. The
algorithm itself is quite short, and relies on simple operations to issue new
random numbers. The parameters chosen for MRG32k3a are such that it has a full
period of $2^{191}$ numbers. This period is fairly enough since
L'Ecuyer suggests that periods between $2^{100}$ and
$2^{200}$ are highly sufficient even for large-scale simulations.
MRG32k3a has been designed to produce independent streams and sub-streams from
its original random sequence thanks to its parameters that enable safe Sequence
Splitting \cite{Hill.etal.2012}. The internal parameters split the initial sequence
into $2^{64}$ adjacent streams of $2^{127}$ random numbers,
themselves divided into sub-streams containing $2^{76}$ elements.

Now considering the distribution aspect, we can assign a stream or a sub-stream
to each computational element according to the Sequence Splitting technique. As
long as we are focusing on parallel applications that are CUDA-enabled, we are
dealing with fine-grained Single Instruction, Multiple Threads (SIMT)
applications. It means that the computational elements are, in our case, the
logical threads of a CUDA kernel and the principle of SIMT is to load the
device with as much threads as possible. Still, we do not expect having to deal
with more than $2^{64}$ parallel threads, which is the total number of
independent streams bearing $2^{127}$ random numbers each that MRG32k3a
can provide.

\subsection{TinyMT}

TinyMT is the latest offspring from the Mersenne Twister family. TinyMT is not
described in any scientific article yet, but information about it can be found
on its dedicated webpage \cite{Saito2011}. This PRNG matches the requirements we
have formulated for a PRNG to be integrated into Shoverand: it is described as producing a good quality output, according to
TestU01 statistical tests, and displays a long-enough period of
$2^{127}$ numbers. TinyMT leverages parameterization to provide
highly independent streams, each stream being represented by a unique
parameterized status.

The theoretical aspect of this approach is very satisfying, but TinyMT
parameterized statuses need to be precomputed by a piece of software called
Dynamic Creator (DC), which is shipped with the PRNG as an open-source binary.
The idea here is to initialize each computing elements with a different status,
since DC can create over $2^{32} × 2^{16}$ independent statuses.
However, memory footprint considerations forced us to propose a hybrid
implementation where the same independent parameterized status is shared among
all the threads of a CUDA block. Independence between random sources is
achieved by feeding each thread with a sub-stream of the original stream,
following the Sequence Splitting technique. To do so, the original stream is
sliced in equal chunks whose starting point, the seed status, is indicated to
threads depending on their identifier. As a consequence, each thread will
always consume the same random sequence in different replications of the same
execution, thus ensuring reproducibility of the experience.

\subsection{Philox and Threefry}

Philox and Threefry are counter-based PRNGs \cite{Salmon.etal.2011} also relying on
parameterization to solve random streams partition concerns. Like any other
PRNG considered in this study, they are Crush-resistant and display good
performance in regards to their low memory footprint and high numbers
throughput. They appear to be better suited than TinyMT (or any other member of
the Mersenne Twister family) to a straightforward GPU implementation since
their parameters are formed by a single key that can be set at runtime
according to each thread’s unique identifier. Please note that the GPU
implementation of these PRNGs is directly provided by their authors. Both CUDA
and OpenCL implementations are proposed for Philox and Threefry.

\section{Case study: generating pseudorandom numbers in a CUDA kernel with Shoverand}
\label{case_study_user}
In this section, we describe a major aspect of Shoverand: its user-friendly
interface. We will see that on both host and device sides, our API is very
expressive while remaining quite concise. Shoverand competes with two major
counterparts in the CUDA world, both coming from an NVIDIA initiative, named
Thrust \cite{Hoberock.Bell2010} and cuRand \cite{NVIDIA2012}. These two libraries are
also providing random number generation features but vary from Shoverand on
several points that we will compare in this section.

\subsection{Host side: Initialization phase}

From the end-user's point of view, Shoverand requires an
initialization phase in order to allocate its internal data structures on the
device and perform some initializations. As a matter of fact, depending on the
chosen PRNG, initialization might involve external data to be read from a
parameter file, as is the case with TinyMT for instance.

We previously saw that we needed to consider the distribution technique and the
PRNG algorithm as a pair. As a consequence, distribution techniques vary from
one PRNG to another in Shoverand, but their initialization phase require the
same data, which is basically the number of CUDA blocks that the kernel using
the PRNG will spawn. This data being stored prior to the kernel call, no
superfluous parameter needs to be passed to the kernel. This feature allows
users not to have their hands tied when designing their kernels, since
Shoverand does not impact kernels' prototypes like other
libraries do.

As a result, the host side initialization phase boils down to a single call to a
static method named init, which must be provided by every PRNG implementation
to satisfy Shoverand's rules.

\subsection{Device side: Computation phase}

Using the device side of Shoverand is even simpler than the host side. You only
have to create an instance of the PRNG you want to use and let its
class' constructor do the rest. Device side initializations
are performed behind the scenes by the constructor so that users have nothing
to do. Then, you can pick random numbers by calling the next() method on the
previously created object. If you are used to random number generation
facilities offered by high-level languages such as Java, making use of
Shoverand in your kernel is really intuitive.

\subsection{Comparison with Thrust and cuRand}

Thrust and cuRand are two projects developed by NVIDIA fellows. While cuRand is
part of the CUDA SDK, Thrust is an external open-source library that can be
downloaded from an Internet repository. In the paper originally introducing
Shoverand \cite{Passerat-Palmbach.etal.2011}, we had surveyed these two libraries
and identified their major drawbacks that led us to design Shoverand. The
following lines investigate the changes brought by the state-of-the-art
versions of Thrust and cuRand.

cuRand is NVIDIA’s solution to random number generation on GPU. In
\cite{Passerat-Palmbach.etal.2011}, we mentioned that cuRand suffered from the
poor statistical quality of the PRNGs it embedded. The last version of this
library partially solves this problem by following the advice we made in our
previous paper. Now, cuRand embeds renowned high-quality PRNGs such as MRG32k3a
\cite{LEcuyer1999} and MTGP \cite{Saito.Matsumoto2012}, whose pseudorandom streams are
stated as “Crush-resistant”.

On the other hand, cuRand's API remains poor and forces users
to add extra parameters in the prototypes of every kernel taking advantage of
the library. The C API is not generic and loses its consistency when
non-default options are enabled: for instance, MTGP's
initialization step is achieved through a dedicated call in cuRand that is
totally irrelevant when used with another PRNG. This approach is not convenient
when you want to quickly swap PRNGs to study the impact of various random
sources on a given application.

Thrust is an open source GP-GPU-enabled general purpose library developed by
NVIDIA fellows. This project aims at providing a GP-GPU-enabled library
equivalent to some classic general-purpose C++ libraries, like STL or Boost.
Classes are split through several namespaces, such as Thrust::random. The
latter contains all classes and methods related to random numbers generation on
GP-GPU. Thrust::random implements three PRNGs, each through a different C++
class template. We find a Linear Congruential Generator (LCG), a Linear
Feedback Shift (LFS) and a Subtract With Carry (SWC), which are widely reckoned
as not adapted to High Performance Computing.

Still, Thrust offers a nice API that mirrors Boost's. The
random namespace provides user-friendly features very close to
Shoverand's abilities. For instance, neither explicit
initializations nor parameters are required in order to benefit from random
number generation facilities in a kernel.

As a conclusion, we have on the one hand cuRand, a library that is improving
after having been criticized by the research community for the statistical
characteristics of its embedded PRNGs, and that exposes a restrictive API; and
on the other hand, we have Thrust and its nice “à la Boost” API, yet powered by
poor quality PRNGs. Then, Shoverand is based upon the good achievements from
these two libraries: indeed, it exposes a user-friendly API while integrating
only Crush-resistant PRNGs.

Listing \ref{listing_shoverand} shows off how to pick up pseudorandom numbers from Shoverand.
In this code snippet, we consider both the initialization on the host side and the random number
generation on the device side:

\begin{lstlisting}[label=listing_shoverand, caption={Example of use of Shoverand's API}]
using shoverand::RNG;
using shoverand::MRG32k3a;

// kernel using Shoverand
__global__ void fooKernel(float* ddata) {

        RNG < float, MRG32k3a >         rng;

        ddata[blockDim.x * blockIdx.x + threadIdx.x] = rng.next();
}

...

RNG< float, MRG32k3a >::init(block_num);

fooKernel <<< block_num, thread_num >>>(d_data);

RNG< float, MRG32k3a >::release();
\end{lstlisting}

\section{Case study: embedding a new PRNG into Shoverand}
\label{case_study_developer}
Shoverand is not only a library but also a framework that allows PRNG developers
to insert their own proposals as long as they follow some rules. In order
to help them in their task, Shoverand employs a mechanism called concept
checking that lets us express constraints in Shoverand's code
that will be checked at compile-time for all the classes that inherit from
ours. Concept checking is a generic programming feature available through
different implementations with C++. It was introduced by \cite{Siek.Lumsdaine2000}
and implemented in the Boost Concept Check Library (BCCL). The application
scope of concept checking is quite wide and goes from interface checking (i.e.,
verifying the presence of a given method within a class), to assessing the
presence of a particular member from a given type in a class.

Such a mechanism forces developers to match our interface without having to
introduce costly runtime techniques like polymorphism and consequently virtual
methods.

Thanks to the BCCL, we are able to design constraints that will make any
compilation attempt fail if they are not met. In Shoverand, we force every PRNG
algorithm class to inherit from our base RNG class. Then, each user-defined
subclass must define at least 3 methods: init() and release(), which deal with
parameters allocation and initialization from the host side, and next(), which
picks up the next random number within a kernel, on the device side. In the
same way, BCCL also permits us to verify that developers have provided two
members to their class: the seed and parameterized statuses. These two members
respectively represent the current state of the PRNG and its initial
parameters. Fig. \ref{prng_uml} sketches a UML class diagram of the expected content of a
PRNG class to be embedded in Shoverand:

\begin{figure}[!h]
\centering
\includegraphics[keepaspectratio, scale=0.5]{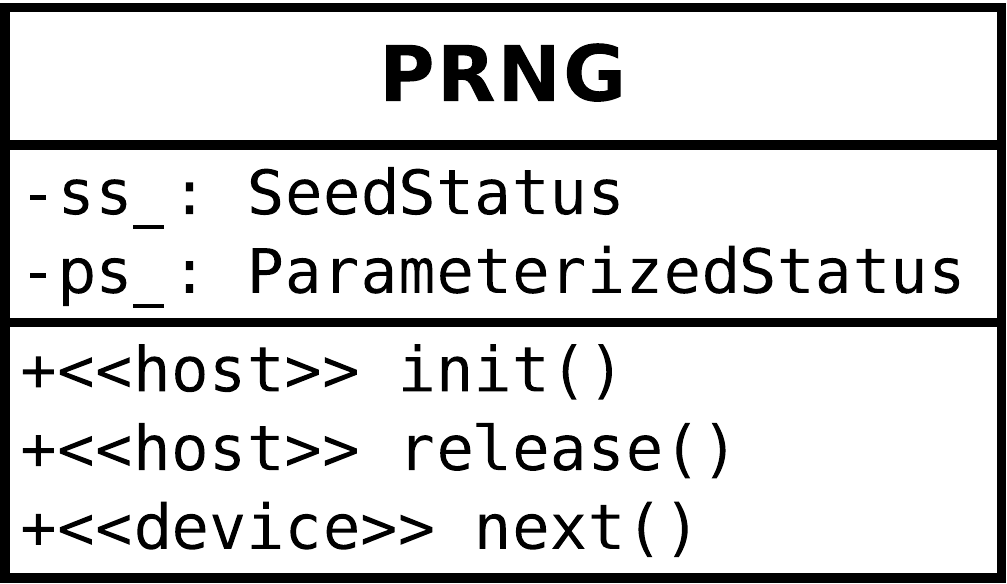}
\caption{UML class diagram of the expected interface of a PRNG in Shoverand}
\label{prng_uml}
\end{figure}

\section{Conclusion}
In this paper, we demonstrate how parallel stochastic simulations can be seriously impacted by the quality of the
underlying Random Number Generators (RNGs), and the partitioning techniques used
to feed such applications \cite{Hill.etal.2012}. Still, we have at our disposal a set of
very reliable sequential generators and a set of distribution techniques
adapted to the different kinds of generators.

From a practitioner's point of view, we have described Shoverand: a CUDA library that embeds the PRNGs displaying
the best statistical quality. Moreover, its user-friendly interface make it a very serious choice when faced to its counterparts.

Shoverand is freely available for download at the following URL: \url{http://http://forge.clermont-universite.fr/projects/shoverand}.

\section*{Acknowledgements}
The authors would like to thank Luc Touraille for his careful reading and useful suggestions on the draft. This work received financial support from the Auvergne Regional Council.

\bibliographystyle{plain}
\bibliography{hpcs2012_tutorial}

\end{document}